\documentclass[a4paper]{article}
\usepackage{layaureo}

\usepackage{xspace,url}
\usepackage{paralist}
\usepackage{dcolumn}
\usepackage{booktabs}
\usepackage{dcolumn}
\usepackage{mathtools,amssymb,amsmath}

\usepackage{algpseudocode}
\usepackage{algorithm}
\usepackage{multicol}
\usepackage{tikz}
\pgfdeclarelayer{edgelayer}
\pgfdeclarelayer{nodelayer}
\usetikzlibrary{arrows, shapes.geometric,snakes}
\usepackage{tkz-euclide}
\usepackage[T1]{fontenc}
\usepackage{caption}
\usepackage{multicol}
\usepackage{multirow}
\usepackage{amssymb,bbm,stmaryrd,array,paralist,delarray}
\usepackage[font=footnotesize]{subcaption}
\usepackage[font=small,labelfont=bf,tableposition=top]{caption}

\usepackage{pgfplots}
\usepackage{pgfplotstable}
\usepackage{tikz}
\pgfdeclarelayer{edgelayer}
\pgfdeclarelayer{nodelayer}
\usepackage{tkz-euclide}	
\usetikzlibrary{arrows, shapes.geometric,snakes}

\usepackage{subfig}
\newcommand{\plotScale}{0.67}
\newcommand{\plotHeight}{6.1cm}

\newcommand{\sol}{SOLOIST\xspace}
\newcommand{\mr}{MapReduce\xspace}

\usepackage{mathptmx}

\makeatletter
\renewcommand{\ALG@beginalgorithmic}{\scriptsize}
\makeatother

\usepackage[scaled=0.8]{DejaVuSansMono}

\begin{document}

\title{Trace checking of Metric Temporal Logic with Aggregating Modalities using MapReduce}

\author{
Domenico Bianculli\\
SnT Centre - University of Luxembourg, Luxembourg\\
domenico.bianculli@uni.lu\\
\and
Carlo Ghezzi, 
Sr\dj{}an Krsti\'c\\
DEEP-SE group - DEIB - Politecnico di Milano, Italy\\
\{carlo.ghezzi,srdjan.krstic\}@polimi.it\\
}

\maketitle
\begin{abstract}
  Modern complex software systems produce a large amount of execution
  data, often stored in logs. These logs can be analyzed using trace
  checking techniques to check whether the system complies with its
  requirements specifications. Often these specifications express
  quantitative properties of the system, which include timing
  constraints as well as higher-level constraints on the occurrences
  of significant events, expressed using aggregate operators.

In this paper we present an algorithm that exploits the MapReduce 
programming model to check specifications expressed in a metric temporal logic with aggregating modalities, over large execution traces. The algorithm exploits the structure of the formula to parallelize the evaluation, with a significant gain in time. We report on the assessment of the implementation---based on the Hadoop framework---of the proposed algorithm and comment on its scalability.

\end{abstract}


\section{Introduction}
\label{sec:intro}
Modern software systems, such as service-based applications (SBAs), are
built according to a modular and decentralized architecture, and
executed in a distributed environment. Their development and their
operation depend  on many stakeholders, including the providers of
various third-party services and the integrators that realize
composite applications by orchestrating third-party services. Service
integrators are responsible to the end-users for guaranteeing an
adequate level of quality of service, both in terms of functional and
non-functional requirements. This new type of software has triggered
several research efforts that focus on the specification and 
verification of SBAs.

In previous work~\cite{bgps:icse2012}, some of the authors presented
the results of a field study on property specification
patterns~\cite{dwyer1998:property-specif} used in the context of
SBAs, both in industrial and in research
settings. The study identified a set of property specification
patterns specific to service provisioning. Most of these patterns are
characterized by the presence of aggregate operations on sequences of
events occurring in a given time window, such as ``the average
distance between pairs of events (e.g., average response time)'',
``the number of events in a given time window'', ``the average (or
maximum) number of events in a certain time interval over a certain
time window''. This study led to the definition of  \sol~\cite{bianculli13:_tale_solois} 
(\emph{SpecificatiOn Language fOr servIce compoSitions inTeractions}),
a metric temporal logic with new temporal modalities that support aggregate operations on events
occurring in a given time window. The new temporal modalities capture,
in a concise way, the new property specification patterns presented
in~\cite{bgps:icse2012}.

\sol has been used in the context of \emph{offline trace checking} of
service execution traces. Trace checking (also called \emph{trace
  validation}~\cite{mrad13:_babel} or \emph{history
  checking}~\cite{Felder:1994:VRS:201024.201034}) is a procedure for
evaluating a formal specification over a log of recorded events
produced by a system, i.e., over a temporal evolution of the system.
Traces can be produced at run time by a proper monitoring/logging
infrastructure, and made available at the end of the service execution
to perform offline trace checking. We have proposed
procedures~\cite{bbgks-fase2014,bgks-TR2013} for offline checking of
service execution traces against requirements specifications written
in \sol using bounded satisfiability checking
techniques~\cite{pradella2013:bounded-satisfi}. Each of the procedures
has been tailored to specific types of traces, depending on the
degree of sparseness of the trace (i.e., the ratio between the number
of time instants where significant events occur and those in which they do not). 
The procedure described in~\cite{bbgks-fase2014} is optimized
for sparse traces, while the one presented in~\cite{bgks-TR2013} is
more efficient for dense traces.

Despite these optimizations,
our experimental evaluation revealed, in both procedures, an intrinsic
limitation in their scalability. This limitation is determined by the
size of the trace, which can quickly lead to memory saturation. 
This is a very common problem, because execution traces can
easily get very large, depending on the running time captured
by the log, the systems the log refers to (e.g., several virtual
machines running on a cloud-based infrastructure), and the types of
events recorded. For example, granularity can range from high-level
events (e.g., sending or receiving messages) to low-level events
(e.g., invoking a method on an object).  Most log analyzers that
process data streams~\cite{Cugola:2012:CEP:2221990.2222305} or perform
data mining~\cite{prom-xes-verbeek2011} only partially
solve the problem of checking an event trace against requirements 
specifications, because of the limited expressiveness of the
specification language they support. Indeed, the analysis of a trace may
require checking for complex properties, which can refer to specific
sequence of events, conditioned by the occurrence of other event
sequence(s), possibly with additional constraints on the distance
among events, on the number of occurrences of events, and on various
aggregate values (e.g., average response time).
\sol addresses these limitations as we discussed above.

The recent advent of cloud computing has made it possible to process
large amount of data on networked commodity hardware, using a
distributed model of computation. One of the most prominent
programming models for distributed, parallel computing is
\emph{\mr}~\cite{Dean:2008:MSD:1327452.1327492}.  The \mr model allows
developers to process large amount of data by breaking up the analysis
into independent tasks, and performing them in parallel on the various
nodes of a distributed network infrastructure, while exploiting, at
the same time, the locality of the data to reduce unnecessary
transmission over the network. However, porting a
traditionally-sequential algorithm (like trace checking) into a
parallel version that takes advantage of a distributed computation
model like \mr is a non-trivial task.

The main contribution of this paper is an algorithm that exploits the
\mr programming model to check large execution traces against
requirements specifications written in \sol. The algorithm exploits
the structure of a \sol formula to parallelize its evaluation, with
significant gain in time. We have implemented the algorithm in Java using the
Apache Hadoop framework~\cite{foundation07:_hadoop_mapred}. We have
evaluated the approach in terms of its scalability and with respect to
the state of art for trace checking of LTL properties using \mr~\cite{DBLP:conf/rv/BarreKSOH12}.

The rest of the paper is structured as follows.  First we provide some
background information, introducing \sol in Sect.~\ref{sec:sol} and
then the \mr programming model in Sect.~\ref{sec:mr}.
Section~\ref{sec:alg} presents the main contribution of the paper,
describing the algorithm for trace checking of \sol properties using
the \mr programming model.
 Section~\ref{sec:rel} discusses related
work. Section~\ref{sec:eval} presents the
evaluation of the approach, both in terms of scalability and in terms of a
comparison with the state of the art for \mr-based trace checking of
temporal properties. Section~\ref{sec:conc} provides some concluding remarks.


\section{\sol}
\label{sec:sol}

In this section we provide a brief overview of \sol; for the
rationale behind the language and a detailed explanation of its
semantics see~\cite{bianculli13:_tale_solois}.

The syntax of \sol is defined by the following grammar:
$\phi \Coloneqq p \mid  \neg \phi \mid \phi \land
\phi \mid \phi \mathsf{U}_I \phi \mid \phi
\mathsf{S}_I \phi \mid \mathfrak{C}^{K}_{\bowtie n}(\phi) \mid
\mathfrak{U}^{K,h}_{\bowtie n}(\phi) \mid \mathfrak{M}^{K,h}_{\bowtie
  n}(\phi) \mid \mathfrak{D}_{\bowtie n}^K
  (\phi,\phi)$, 
where $p \in \Pi$, with $\Pi$ being a finite set of atoms. In practice, we use atoms 
to represent different events of the trace. $I$  is a
nonempty interval over $\mathbb{N}$;  $\bowtie \  \in \{<,\leq,\geq,>,=\}$;  $n,K,h$
range over $\mathbb{N}$. 
Moreover, for the $\mathfrak{D}$ modality, we
require that the subformulae pair $(\phi, \psi)$ evaluate to true in
alternation.

The $\mathsf{U}_I$ and $\mathsf{S}_I$ modalities are, respectively,
the metric ``\emph{Until}'' and ``\emph{Since}'' operators. Additional
temporal modalities can be derived using the usual conventions; for
example ``\emph{Next}'' is defined as $\mathsf{X}_I \phi \equiv \bot \mathsf{U}_I \phi$; 
``\emph{Eventually in the Future}'' as $\mathsf{F}_I \phi \equiv \top \mathsf{U}_I \phi$
and ``\emph{Always}'' as $\mathsf{G}_I \phi \equiv \neg (\mathsf{F}_I \neg \phi)$, 
where $\top$ means ``true'' and $\bot$ means ``false''. Their 
past counterparts can be defined using ``Since'' modality in a similar
way.
The remaining modalities are called \emph{aggregate} 
modalities and are used to express the property specification patterns
characterized in~\cite{bgps:icse2012}. The $\mathfrak{C}^{K}_{\bowtie n}(\phi)$ modality
states a bound (represented by $\bowtie n$) on the number of occurrences 
of an event $\phi$ in the previous $K$ time instants; it is also
called the ``\emph{counting}'' modality.
The $\mathfrak{U}^{K,h}_{\bowtie n}(\phi)$ (respectively,
$\mathfrak{M}^{K,h}_{\bowtie n}(\phi)$) modality expresses a bound on
the average (respectively, maximum) number of occurrences of an event
$\phi$, aggregated over the set of right-aligned adjacent
non-overlapping subintervals within a time window $K$; it can express
properties like ``the average/maximum number
of events per hour in the last ten hours''.  A subtle difference in the
semantics of the $\mathfrak{U}$ and $\mathfrak{M}$ modalities is that
$\mathfrak{M}$ considers events in the (possibly empty) tail interval,
i.e., the leftmost observation subinterval whose length is less than $h$,
while the $\mathfrak{U}$ modality ignores them.  The
$\mathfrak{D}^{K}_{\bowtie n}(\phi,\psi)$ modality expresses a bound
on the average time elapsed between occurrences of pairs of specific adjacent events
$\phi$ and $\psi$ in the previous $K$ time instants; it can
be used to express properties like the average response time of a service.

\begin{figure*}[tb]
\begin{center}
\begin{scriptsize}
\begin{tabular}{>{$}p{40mm}<{$}c>{$}l<{$}}
   (w, i)  \models
   p &{if{f}}& p \in \sigma_i\\
   (w, i)  \models \neg \phi
   &{if{f}}&  (w, i) \not
   \models \phi \\
   (w, i)  \models \phi \land \psi
   &{if{f}}& (w, i) 
   \models \phi \land (w, i)
   \models \psi\\
   (w, i)  \models \phi \mathsf{S}_I
   \psi &{if{f}}& \text{for some } j < i, \tau_i - \tau_j \in
   I,  (w,j)  \models \psi
    \text{ and for all } k, j < k < i,  (w, k)  
   \models \phi\\
   (w, i)  \models \phi \mathsf{U}_I
   \psi &{if{f}}& \text{for some } j > i, \tau_j - \tau_i \in
   I,  (w,j)  \models \psi
    \text{ and for all }k, i < k < j,  (w, k)  
   \models \phi\\
   (w, i)  \models
   \mathfrak{C}_{\bowtie n}^{K}(\phi) &{if{f}}& 
   c(\tau_i - K, \tau_i,\phi) \bowtie n \text{ and } \tau_i \geq K\\
   (w, i)  \models
   \mathfrak{U}_{\bowtie n}^{K,h}(\phi) &{if{f}}&
   \dfrac{ c(\tau_i - \lfloor \frac{K}{h}\rfloor h, \tau_i,\phi)}{\lfloor \frac{K}{h} \rfloor} \bowtie n \text{ and } \tau_i \geq K
    \\
   (w, i)  \models
   \mathfrak{M}_{\bowtie n}^{K,h}(\phi) &{if{f}}& 
  \max\left\{\bigcup_{m=0}^{\left \lfloor \frac{K}{h} \right \rfloor}
     \left\{ c(\mathit{lb}(m),\mathit{rb}(m),\phi) \right\} \right\}
   \bowtie n \text{ and } \tau_i \geq K\\
   (w, i)  \models \mathfrak{D}_{\bowtie n}^K
   (\phi,\psi)  &{if{f}}&
   \dfrac{\sum_{(s,t) \in d(\phi,\psi,\tau_i,K)}
     (\tau_t-\tau_s)}{|d(\phi,\psi,\tau_i,K)|}
   \bowtie n  \text{ and } \tau_i \geq K\\[5mm]
   \multicolumn{3}{l}{where
$c(\tau_a,\tau_b,\phi) = | \left \{ s \mid
\tau_a < \tau_s \leq \tau_b  \text{ and } (w, s) \models \phi \right
\} |$, $\mathit{lb}(m)=\max\{\tau_i - K,  \tau_i - (m+1) h \} $,
$\mathit{rb}(m)=\tau_i - mh$, and}\\
\multicolumn{3}{l}{$d(\phi,\psi,\tau_i, K) = \left\{(s,t) \mid \tau_i - K <
  \tau_s \leq \tau_i \text{ and } (w, s) 
  \models \phi, t= \min\{ u \mid \tau_s < \tau_u \leq
  \tau_i, (w,
  u)  \models \psi\} \right\}$}
\end{tabular}


\end{scriptsize}
\end{center}
\caption{Formal semantics of \sol}
\label{fig:semantics}\end{figure*}

The formal semantics of \sol is 
 defined
on timed $\omega$-words~\cite{Alur:1994:TTA:180782.180519} over
$2^{\Pi} \times \mathbb{N}$.
A timed sequence $\tau = \tau_0 \tau_1 \ldots $ is an infinite sequence of values $\tau_i \in \mathbb{N}$ with $\tau_i > 0$
satisfying $\tau_i<\tau_{i+1}$, for all $i\geq 0$, i.e., the sequence
increases strictly monotonically.
A timed $\omega$-word over alphabet $2^{\Pi}$ is a pair $(\sigma, \tau)$ where $\sigma= \sigma_0 \sigma_1 \ldots$ is an infinite word
over $2^{\Pi}$ and $\tau$ is a timed sequence. A timed language over $2^{\Pi}$ is a set of timed words over the same alphabet.  
Notice that there is a distinction between the integer position $i$ in the timed $\omega$-word and the corresponding timestamp $\tau_i$.
Figure~\ref{fig:semantics} defines the satisfiability relation $(w, i) \models \phi$ 
for every timed $\omega$-word $w$, every position $i\ge 0$ and for every \sol formula $\phi$.
For the sake of simplicity, hereafter we express the $\mathfrak{U}$
modality in terms of the  $\mathfrak{C}$ one, based on this
definition:  $\mathfrak{U}_{\bowtie n}^{K,h} (\phi) \equiv
\mathfrak{C}_{\bowtie n \cdot \lfloor \frac{K}{h} \rfloor}^{\lfloor \frac{K}{h} \rfloor \cdot h} (\phi) 
$, which can be derived from the semantics in Fig.~\ref{fig:semantics}.

We remark that the version of \sol presented here is a restriction of the original one introduced
in~\cite{bianculli13:_tale_solois}: to simplify the presentation in the
next sections, we dropped first-order quantification on finite domains and limited the
argument of the $\mathfrak{D}$ modality to only one pair of
events; as detailed in~\cite{bianculli13:_tale_solois}, these
assumptions do not affect the expressiveness of the language.

\sol can be used to express some of the most common specifications
found in service-level agreements (SLAs) of SBAs. For example the property:
``The average response time of operation \texttt{A} is always less
than 5 seconds within any 900 second time window, before operation \texttt{B} is invoked'' can be expressed as: 
$\mathsf{G}(\text{\texttt{B}}_\mathit{start} \rightarrow \mathfrak{D}^{900}_{< 5}(\text{\texttt{A}}_{\mathit{start}}, \text{\texttt{A}}_{\mathit{end}}))$,
where \texttt{A} and \texttt{B} correspond to generic service invocations and 
each operation has a \emph{start} and an \emph{end} event, denoted with the 
corresponding subscripts.

We now introduce some basic concepts that will be used in the
presentation of our distributed trace checking algorithm in
Sect.~\ref{sec:alg}.  Let $\phi$ and $\psi$ be \sol formulae.  We
denote with $\mathsf{sub}(\phi)$ the set of all subformulae of
$\phi$; notice that for
\emph{atomic} formulae $a \in \Pi, \ \mathsf{sub}(a)=\emptyset$.  
The set of \emph{atomic} subformulae (or \emph{atoms}) of formula
$\phi$ is defined as $\mathsf{sub}_a(\phi)=\{a \mid a \in
\mathsf{sub}(\phi), \ \mathsf{sub}(a)=\emptyset\}$.
The set $\mathsf{sub}_d(\phi)=\{\alpha \mid \alpha \in
\mathsf{sub}(\phi), \forall \beta \in \mathsf{sub}(\phi), \alpha
\notin \mathsf{sub}(\beta)\}$ represents the set of all \emph{direct
  subformulae} of $\phi$; $\phi$ is called the \emph{superformula} of
all formulae in $\mathsf{sub}_d(\phi)$.
The notation $\mathsf{sup}_{\psi}(\phi)$ denotes the set of all
subformulae of $\psi$ that have formula $\phi$ as \emph{direct
  subformula}, i.e., $\mathsf{sup}_{\psi}(\phi)=\{\alpha \mid \alpha
\in \mathsf{sub}(\psi), \phi \in \mathsf{sub}_d(\alpha) \}$.
The subformulae in $\mathsf{sub}(\psi)$ of a formula $\psi$ form a
lattice with respect to the partial ordering induced by the inclusion
in sets $\mathsf{sup}_{\psi}(\cdot)$ and $\mathsf{sub}_d(\cdot)$, with
$\psi$ and $\emptyset$ being the \emph{top} and \emph{bottom} elements
of the lattice, respectively.
We also introduce the notion of the \emph{height} of a \sol formula,
which is defined recursively as:
\[ h(\phi) = \left\{ 
  \begin{array}{l l}
    \mathsf{max}\{h(\psi) \mid \psi \in \mathsf{sub}_d(\phi)\}+1 & \quad \text{if }\mathsf{sub}_d(\phi)\neq\emptyset\\
    0 & \quad \text{otherwise.}
  \end{array} \right.\]
\\
We exemplify these concepts using formula $\gamma\equiv \mathfrak{C}^{40}_{\bowtie 3}(a \land b) \mathsf{U}_{(30,100)} \neg
c$.\\ Hence $\mathsf{sub}(\gamma)=\{a,b,c, a\land b, \neg c,
\mathfrak{C}^{40}_{\bowtie 3}(a \land b)\}$ is the set of all subformulae of 
$\gamma$; $\mathsf{sub}_a(\gamma)=\{a,b,c\}$ is the set of \emph{atoms} in $\gamma$;
$\mathsf{sub}_d(\gamma)=\{\mathfrak{C}^{40}_{\bowtie 3}(a \land b),\neg c\}$ is the set 
of direct subformulae of $\gamma$; $\mathsf{sup}_{\gamma}(a)=\mathsf{sup}_{\gamma}(b)=\{a\land b\}$
shows that the sets of superformulae of $a$ and $b$ in $\gamma$ coincide; and the height of $\gamma$
is $3$, since $h(a)=h(b)=h(c)=0$, $h(\neg c)=h(a\land b)=1$, $h(\mathfrak{C}^{40}_{\bowtie 3}(a \land b))=2$
and therefore $h(\gamma)=\mathsf{max}\{h(\mathfrak{C}^{40}_{\bowtie 3}(a \land b)),h(\neg c)\}+1=3$.


\section{The \mr programming model}
\label{sec:mr}

\mr~\cite{Dean:2008:MSD:1327452.1327492} is a programming model for
processing and analyzing large data sets using a parallel, distributed
infrastructure (generically called ``cluster''). At the basis of the \mr abstraction there are two
functions, \textit{map} and \textit{reduce}, that are inspired by (but
conceptually different from) the homonymous functions that are
typically found in functional programming languages.
The \emph{map} and \emph{reduce} functions are defined
by the user; their signatures are \texttt{map(k1,v1) $\rightarrow$
  list(k2,v2)} and \texttt{reduce(k2,list(v2)) $\rightarrow$
  list(v2)}.
 The idea of \mr is to apply a
 \emph{map} function to each logical entity in the input (represented
 by a key/value pair)
in order to compute a set of intermediate key/value pairs, and then
applying a \emph{reduce} function to all the values that have the same
key in order to combine the derived data appropriately. 

Let us illustrate this model with an example that counts the number of
occurrences of each word in a large collection of documents; the
pseudocode is:\\[-20pt]
\begin{multicols}{2}
\begin{scriptsize}
\begin{verbatim}
map(String key, String value)
 //key: document name
 //value: document contents
 for each word w in value:
     EmitIntermediate(w,"1")


reduce(String key, Iterator values):
 //key: a word
 //values: a list of counts
 int result = 0
 for each v in values:
     result += ParseInt(v)
 Emit(AsString(result)
\end{verbatim}
\end{scriptsize}
\end{multicols}
\vspace{-10pt}
The \emph{map} function emits list of pairs, each composed of a word and its
associated count of occurrences (which is just 1). All emitted pairs are
partitioned into groups and sorted according to their key for the
reduction phase; in the example, pairs are grouped and sorted
according to the word they contain. The \emph{reduce} function sums
all the counts (using an iterator to go through the list of counts)
emitted for each particular word (i.e., each unique key).

Besides the actual programming model, \mr brings in a framework that
provides, in a transparent way to developers, parallelization, fault
tolerance, locality optimization, and load balancing. The \mr
framework is responsible for partitioning the input data, scheduling
and executing the \emph{Map} and \emph{Reduce} tasks (also called
\emph{mappers} and \emph{reducers}, respectively) on the machines
available in the cluster, and for managing the communication and the
data transfer among them (usually leveraging a distributed file
system).

More in detail, the execution of a \mr operation (called \emph{job})
proceeds as follows.  First, the framework divides the input into
splits of a certain size using an \emph{InputReader}, generating
key/value $(k,v)$ pairs. It then assigns each input split to
Map tasks, which are processed in parallel by the nodes in the
cluster. A Map task reads the corresponding input split and
passes the set of key/value pairs to the \emph{map} function, which
generates a set of \emph{intermediate} key/value pairs
$(k',v')$. Notice that each run of the \emph{map} function is
stateless, i.e., the transformation of a single key/value pair does
not depend on any other key/value pair.
The next phase is called \emph{shuffle and sort}: it takes the
intermediate data generated by each Map task, sorts them based on the
intermediate data generated from other nodes, divides these data into
regions to be processed by Reduce tasks, and distributes these data on
the nodes where the Reduce tasks will be executed. The division of
intermediate data into regions is done by a \emph{partitioning
  function}, which depends on the (user-specified)
number of Reduce tasks and the key of the intermediate data. Each
Reduce task executes the \emph{reduce} function, 
which takes an intermediate key $k'$ and a set of values 
associated with that key to produce the output data. This output is
appended to a final output file for this reduce partition. The output
of the \mr job will then be available in several files, one for each
Reduce task used.


\section{Trace checking with \mr}
\label{sec:alg}

Our algorithm for trace checking of \sol properties takes as input a
non-empty execution trace $T$ and the \sol formula $\Phi$ to be
checked. The trace $T$ is finite and can be seen as a time-stamped
sequence of $H$ elements, i.e., $T=(p_1,p_2,\ldots,p_{H})$. Each of
these elements is a triple $p_i=(i,\tau_i,(a_1, \ldots, a_{P_i}))$,
where $i$ is the position within the trace, $\tau_i$ the integer
timestamp, and $(a_1, \ldots, a_{P_i})$ is a list of atoms such that
$a_{j_i} \in \Pi$, for all $j_i \in \{1,...P_i\}, P_i\geq 1$ and for
all $i \in \{1,2,\ldots,H\}$.

The algorithm processes the trace iteratively, through subsequent \mr
passes. The number of \mr iterations is equal to height of the \sol
formula $\Phi$ to be checked.  The $l$-th iteration (with $1<l\leq
h(\Phi)$) of the algorithm receives a set of tuples from the
$(l-1)$-th iteration; these input tuples represent all the positions
where the subformulae of $\Phi$ having height $l-1$ hold.  The $l$-th
iteration then determines all the positions where the subformulae of
$\Phi$ with height $l$ hold.

Each iteration consists of three phases: 1) reading and splitting the
input; 2) (\emph{map}) associating each formula with its superformula;
3) (\emph{reduce}) determining the positions where the superformulae
obtained in the previous step hold, given the positions where their
subformulae hold. We detail each phase in the rest of this section.

\subsection{Input reader}
\label{sec:input-reader}
We assume that before the first iteration of the algorithm the input
trace is available in the distributed file system of the cluster; this
is a realistic assumption since in a distribute setting is possible to
collect logs, as long as there is a total order among the timestamps.
The input reader at the first iteration reads the trace directly,
while in all subsequent iterations input readers read the output 
of the reducers of the previous iteration. 

The input reader component of the \mr framework is able to process the
input trace exploiting some parallelism. Indeed,  the \mr framework
exploits the location information of the different fragments of the trace  to parallelize 
the execution of the input reader. For example, a trace split into $n$ fragments can be processed in parallel using $\min(n,k)$ 
machines, given a cluster with $k$ machines.

  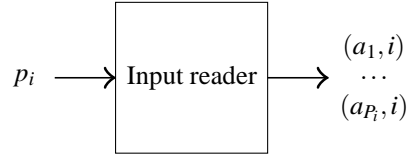
\begin{figure}[t]
      \centering
      \begin{subfigure}{.45\textwidth}
        \centering
        \vspace{0.8cm}
        \begin{algorithmic}
           \Function{Input reader$_{\Phi,k,l}$}{$T_k$}
           \ForAll{$(i,\tau_i,A) \in T_k$} 
           	 \State $TS(i) \gets \tau_i$
          	 \ForAll{$a \in A$}
          	 	\If{$a \in \mathsf{sub}_a(\Phi)$} 
          	  		\State output($a,i$)
          	  	\EndIf
          	 \EndFor
           \EndFor
           \EndFunction
           \end{algorithmic}{\tiny }      
        \caption{Input reader algorithm}\label{fig:reader-alg}
      \end{subfigure}
                  \hfill
      \begin{subfigure}{.45\textwidth}
        \centering
        \begin{tikzpicture}[scale=2]

\def \length{0.4}
\def \heightI{0}
\def \heightU{-0.5}
\def \heightD{0.5}

\def \wI{0}
\def \wII{0.8}
\def \wIII{1.2}
\def \wIV{2.2}
\def \wV{2.6}
\def \wVI{2.9}


\tkzDefPoint(wI+0.6,\heightI+0.7){bla1} 
\tkzDefPoint(wI+0.6,\heightI-1){bla2} 
\tkzDefPoint(wI+0.6,\heightI){I} 

\tkzDefPoint(\wII,\heightI){A1s}
\tkzDefPoint(\wIII,\heightI){A1f}

\tkzDefPoint(\wIII,\heightU){Rs}
\tkzDefPoint(\wIV,\heightD){Rf}

\tkzDefPoint(\wIV,\heightI){A2s}
\tkzDefPoint(\wV,\heightI){A2f}

\tkzDefPoint(wVI,\heightI+0.2){O1} 
\tkzDefPoint(wVI,\heightI){O2} 
\tkzDefPoint(wVI,\heightI-0.2){O3}

\draw (\wIII,\heightU) -- (\wIV,\heightU) -- (\wIV,\heightD) -- (\wIII,\heightD) -- cycle;

\begin{scope}[very thick,decoration={
    markings,
    mark=at position 1 with {\arrow[scale=2]{>}}}
    ] 
    \tkzDrawSegments[postaction={decorate}](A1s,A1f) 

\end{scope}

\begin{scope}[very thick,decoration={
    markings,
    mark=at position 1 with {\arrow[scale=2]{>}}}
    ] 
    \tkzDrawSegments[postaction={decorate}](A2s,A2f)

\end{scope}

\node at (I) {$p_i$};
\node at (O1) {$(a_1,i)$};
\node at (O2) {$\ldots$};
\node at (O3) {$(a_{P_i},i)$};
\node at (1.7,0) {Input reader};

\node at (bla1) {};
\node at (bla2) {};
\normalsize

\end{tikzpicture}
        \caption{Data flow of the Input reader}\label{fig:reader-schema}
      \end{subfigure}
      \caption{Input reader}\label{fig:reader}
    \end{figure}

Figure~\ref{fig:reader-schema} shows how the input reader
transforms the trace at the first iteration: for every atomic
proposition $\phi$ that holds at position $i$ in the original
trace, it outputs a tuple of the form $(\phi,i)$.
The transformation does not happen in the subsequent iterations, since
(as will be shown in Sect.~\ref{sec:reducer})
the output of the reduce phase 
 has the same form $(\phi,i)$.
The
algorithm in Fig.~\ref{fig:reader-alg} shows how input reader handles the $k$-th fragment $T_k$
of the input trace $T$. 
For each time point $i$ and for each atom $p$ that holds in position
$i$ it creates a tuple $(p,i)$. Moreover, for
each time point $i$, it updates a globally-shared associative list of
timestamps $TS$. This list  is used to associate a timestamp
with each time point; its contents are  saved in the distributed file
system, for use during the reduce phase.


\subsection{Mapper}
\label{sec:mapper}
Each tuple generated by an input reader is passed to a mapper at the
local node.  Mappers ``lift'' the formula in the tuple by associating
it with all its superformulae in the input formula $\Phi$.  For
example, given the formula $\Phi \equiv (a \land b) \lor \neg a$, the
tuple $(a,5)$ is associated with formulae $a \land b$ and $\neg a$.
The reduce phase will then exploit the information about the direct
subformulae to determine all the positions in which a superformula
holds.

As shown in Fig.~\ref{fig:mapper}, the output of a mapper are tuples of the form $((\psi,i),(\phi,i))$ where $\phi$ is a direct subformulae of $\psi$ and $i$ is the
position where $\phi$ holds. 
For each received tuple of the form
$(\phi,i)$, the algorithm shown in Fig.~\ref{fig:mapper-alg} loops through all the superformulae $\psi$ of $\phi$ and emits (using the function \emph{output}) a tuple $((\psi,i),(\phi,i))$.

Notice that the key of the intermediate tuples emitted by the mapper
has two parts: this type of key is called a \emph{composite key} and
it is used to perform \emph{secondary sorting} of the intermediate
tuples. Secondary sorting performs the sorting using multiple
criteria, allowing developers to sort not only by the key, but also ``by
value''.  In our case, we perform secondary sorting based on the
position where the subformula holds, in order to decrease the memory
used by the reducer.  To enable secondary sorting, we need to override
the procedure that compares keys, to take into account also the second
element of the composite keys when their first elements are equal. We
have also modified the key grouping procedure to consider only the
first part of the composite key, so that each reducer gets all the
tuples related to exactly one superformula (as encoded in the first
part of the key), sorted in ascending order with respect to the
position where subformulae hold (as encoded in the second part of the
key).

\begin{figure}[t]
      \centering
      \begin{subfigure}{.35\textwidth}
        \centering
        \vspace{0.30cm}
        \begin{algorithmic}
         \Function{Mapper$_{\Phi,l}$}{$(\phi,i)$}
         \If{$l \leq h(\phi)$} 
         \ForAll{$\psi \in \mathsf{sup}_{\Phi}(\phi)$} 
         	\State output($(\psi,i),(\phi,i)$)
         \EndFor
         \EndIf
         \EndFunction
         \end{algorithmic}{\tiny }
         \vspace{0.27cm}
        \caption{Mapper algorithm}\label{fig:mapper-alg}
      \end{subfigure}
                \hspace*{\fill}
      \begin{subfigure}{.6\textwidth}
        \centering
 		\begin{tikzpicture}[scale=2]

\def \length{0.4}
\def \heightI{0}
\def \heightU{-0.5}
\def \heightD{0.5}

\def \wI{0.5}
\def \wII{0.8}
\def \wIII{1.2}
\def \wIV{2.2}
\def \wV{2.6}
\def \wVI{3.1}


\tkzDefPoint(wI+3,\heightI){bla1} 
\tkzDefPoint(wI+0.6,\heightI-0.7){bla2}

\tkzDefPoint(wI,\heightI){I} 

\tkzDefPoint(\wII,\heightI){A1s}
\tkzDefPoint(\wIII,\heightI){A1f}

\tkzDefPoint(\wIII,\heightU){Rs}
\tkzDefPoint(\wIV,\heightD){Rf}

\tkzDefPoint(\wIV,\heightI){A2s}
\tkzDefPoint(\wV,\heightI){A2f}

\tkzDefPoint(wVI,\heightI+0.2){O1} 
\tkzDefPoint(wVI,\heightI){O2} 
\tkzDefPoint(wVI,\heightI-0.2){O3}

\draw (\wIII,\heightU) -- (\wIV,\heightU) -- (\wIV,\heightD) -- (\wIII,\heightD) -- cycle;

\begin{scope}[very thick,decoration={
    markings,
    mark=at position 1 with {\arrow[scale=2]{>}}}
    ] 
    \tkzDrawSegments[postaction={decorate}](A1s,A1f) 

\end{scope}

\begin{scope}[very thick,decoration={
    markings,
    mark=at position 1 with {\arrow[scale=2]{>}}}
    ] 
    \tkzDrawSegments[postaction={decorate}](A2s,A2f)

\end{scope}

\node at (I) {$(\phi,i)$};
\node at (O1) {$((\psi_1, i), (\phi,i))$};
\node at (O2) {$\ldots$};
\node at (O3) {$((\psi_g, i), (\phi,i))$};

\node at (1.7,0) {Mapper};
\node at (bla2) {};

\normalsize

\end{tikzpicture}
        \caption{Data flow of a Mapper}\label{fig:mapper-scema}
      \end{subfigure}
      \caption{Mapper}\label{fig:mapper}
    \end{figure}


\setcounter{subfigure}{0}

\begin{figure}
	\centering
\begin{tabular}{cc}
\begin{minipage}[t]{.5\textwidth}
	\begin{algorithmic}
	\Function{Reducer$_{\mathfrak{D}_{\bowtie n}^{K},\Phi,l,TS}$}{$\mathfrak{D}_{\bowtie n}^{K}(\phi,\psi),\textit{tuples[]}$}
	\If{$h(\mathfrak{D}_{\bowtie n}^{K}(\phi,\psi))=l+1$} 
		\State $p \gets 0,$ $pairs \gets 0,$ $dist \gets 0$
		\ForAll{$(\xi,i) \in \textit{tuples}$}
			\For{$j \gets p+1 \ldots i-1$} 
				\State $\textit{updateDistInterval(j)}$
				\State $\textit{emitDist(j)}$
			\EndFor	
			\If{$\xi=\psi$}
				\State $pairs \gets pairs+1$
				\State $dist \gets dist + (TS(i)-TS(subFmas.last))$
			\EndIf 
			\State $\textit{subFmas.addLast(i)}$
			\State $\textit{updateDistInterval(i)}$
			\State $\textit{emitDist(i)}$
			\State $p \gets i$
		\EndFor	
	\Else
		\ForAll{$(\phi,i) \in \textit{tuples}$} 
	 			\State output($\phi,i$)
		\EndFor
	\EndIf
	\EndFunction
	\end{algorithmic}{\tiny }

\end{minipage} 
&
\begin{minipage}[t]{.5\textwidth}
\begin{algorithmic}
		\Function{Reducer$_{\land,\Phi,l,TS}$}{$\psi,\textit{tuples[]}$}
		\State $p \gets 0,$ $c \gets 1$
		\While{$(\phi,i) \in \textit{tuples}$} 
			\If{$h(\psi)=l+1$} 
				\If{$i=p$} 
					\State $c \gets c+1$
				\Else
					\If{$c=|\mathsf{sub}_d(\psi)|$} 
			 			\State output($\psi,i$)	
					\EndIf
					\State $c \gets 1$
				\EndIf
			\Else
			 	\State output($\phi,i$)	
			\EndIf
			\State $p \gets i$
		\EndWhile
		\EndFunction
		\end{algorithmic}{\tiny }
\end{minipage}\\

\begin{minipage}[t]{.5\textwidth}
\captionof{subfigure}{$\mathfrak{D}$ modality}\label{alg:reducer-dist}
\end{minipage}&

\begin{minipage}[t]{.5\textwidth}
\captionof{subfigure}{Conjunction}\label{alg:reducer-and}		
\end{minipage}\\

\begin{minipage}[t]{.5\textwidth}
		\centering
		\begin{algorithmic}
		\Function{Reducer$_{\mathsf{U_I},\Phi,l,TS}$}{$\phi_1 \mathsf{U_{(a,b)}} \phi_2,\textit{tuples[]}$}
		\If{$h(\phi_1 \mathsf{U_{(a,b)}} \phi_2)=l+1$} 
			\State $p \gets 0$
			\ForAll{$(\xi,i) \in \textit{tuples}$}
				\State $\textit{updateLTLBehavior(i)}$
				\State $\textit{updateMTLBehavior(i)}$
				\If{$\xi = \phi_2$}
					\State $\textit{emitUntil(i)}$
				\EndIf	
				\State $p \gets i$
			\EndFor
		\Else
			\ForAll{$(\phi,i) \in \textit{tuples}$} 
		 			\State output($\phi,i$)	
			\EndFor
		\EndIf
		\EndFunction
		\end{algorithmic}{\tiny }

\end{minipage}&

\begin{minipage}[t]{.5\textwidth}
		\centering
		\begin{algorithmic}
		\Function{Reducer$_{\mathfrak{C}_{\bowtie n}^{K},\Phi,l,TS}$}{$\mathfrak{C}_{\bowtie n}^{K}(\phi),\textit{tuples[]}$}
		\State $p \gets 0,$ $c \gets 0$
		\ForAll{$(\phi,i) \in \textit{tuples}$}
			\State $c \gets c+1$
			\For{$j \gets p+1 \ldots i-1$} 
				\State $\textit{updateCountInterval(j)}$
				\If{$c \bowtie n$}
						\State output($\mathfrak{C}_{\bowtie n}^{K}(\phi),j$)	
				\EndIf
			\EndFor	
			\State $\textit{updateCountInterval(i)}$
			\If{$c \bowtie n$}
				\State output($\mathfrak{C}_{\bowtie n}^{K}(\phi),i$)	
			\EndIf
			\State $p \gets i$
		\EndFor	
		\EndFunction
		\end{algorithmic}{\tiny }

\end{minipage}\\

\begin{minipage}[t]{.5\textwidth}
\captionof{subfigure}{$\mathfrak{U}$ modality}\label{alg:reducer-until}
\end{minipage}&

\begin{minipage}[t]{.5\textwidth}
\captionof{subfigure}{$\mathfrak{C}$ modality}\label{alg:reducer-count}
\end{minipage}\\

\begin{minipage}[t]{.5\textwidth}
		\centering
		\begin{algorithmic}
		\Function{Reducer$_{\neg,\Phi,l,TS}$}{$\neg \phi,\textit{tuples[]}$}
		 \State $p \gets 0$ 
		 \ForAll{$((\phi,i)) \in \textit{tuples}$} 
		 	\For{$j \gets p+1 \ldots i-1$} 
		  		\State output($\neg \phi,j$)
		  	\EndFor
		  	\State $p \gets i$
		 \EndFor
		 \For{$i \gets p+1 \ldots TS.size()$} 
			\State output($\neg \phi,i$)
		 \EndFor
	 	\EndFunction
		\end{algorithmic}{\tiny }

\end{minipage}&

\begin{minipage}[t]{.5\textwidth}
		\centering
		\begin{algorithmic}
		\Function{Reducer$_{\mathfrak{M}_{\bowtie n}^{K,h},\Phi,l,TS}$}{$\mathfrak{M}_{\bowtie n}^{K,h}(\phi),\textit{tuples[]}$}
		\State $p \gets 0$
		\ForAll{$(\xi,i) \in \textit{tuples}$}
			\For{$j \gets p+1 \ldots i-1$} 
				\State $\textit{updateMaxInterval(j)}$
				\State $\textit{emitMax(j)}$
			\EndFor	
			\State $\textit{updateMaxInterval(i)}$
			\State $\textit{emitMax(i)}$
			\State $p \gets i$
		\EndFor	
		\EndFunction
		\end{algorithmic}{\tiny }

\end{minipage}\\

\begin{minipage}[t]{.5\textwidth}
\captionof{subfigure}{Negation}\label{alg:reducer-not}
\end{minipage}&

\begin{minipage}[t]{.5\textwidth}
\captionof{subfigure}{$\mathfrak{M}$ modality}\label{alg:reducer-max}
\end{minipage}\\

\end{tabular}

    \caption{Reduce algorithms}\label{algs:reduce}
\end{figure}

\subsection{Reducer}
\label{sec:reducer}
In the reduce phase, at each iteration $l$, reducers calculate all positions 
where subformulae with height $l$ hold.  The total
number of reducers running in parallel at the $l$-th iteration is the
minimum between the number of subformulae with height $l$ in the input
formula $\Phi$ and the number of machines in the cluster multiplied by
the number of reducers available on each node.  Each reducer calls an
appropriate reduce function depending on the type of formula used as
key in the input tuple.  The initial data shared by all reducers is
the input formula $\Phi$, the index of the current \mr iteration $l$
and the associative map of timestamps $\mathit{TS}$.

In the rest of this section we present the algorithms of the reduce
function defined for \sol connectives and modalities.  For space
reasons we limit the description to the algorithms for negation
($\neg$) and conjunction ($\land$), and for the modalities
$\mathsf{U}_I$, $\mathfrak{C}_{\bowtie n}^{K}$, $\mathfrak{M}_{\bowtie
  n}^{K,h}$, and $\mathfrak{D}_{\bowtie n}^{K}$. The other temporal
modalities can be expressed in a way similar to the \textit{Until}
modality $\mathsf{U}_I$. 
   In the various algorithms we use several
   auxiliary functions whose pseudocode is available
 in the appendix.

\textbf{Negation.}
When the key refers to a negated superformula, the reducer emits a tuple at every position where the subformula does not hold, i.e., at every position that does not occur in the input tuples received from the
mappers. The algorithm in Fig.~\ref{alg:reducer-not} shows how output
tuples are emitted. If no tuples are received then the reducer emits tuples at each position. 
Otherwise, it keeps track of the  position $i$ of the current tuple
and the position $p$ of the previous tuple and emits tuples at
positions $[p+1, i-1]$.

\textbf{Conjunction.}
We extend the binary $\land$ operator defined in Sect.~\ref{sec:sol}
to any positive arity; this extension does not
change the language  but improves  the conciseness of the formulae.
With this extension, conjunction $a \land b \land c$ is represented as
a single conjunction with 3 subformulae and has height equal to 1.
Tuples $(\phi,i)$ received from the mapper may refer to any subformula $\phi$ of a conjunction.

In the algorithm in Fig.~\ref{alg:reducer-and} we process all the tuples sequentially. 
First, we check if the height of each subformula is
consistent with respect to the iteration in which they are
processed. In fact, mappers can emit some tuples before the ``right''
iteration in which they should be processed, since subformule of a conjunction may have different
height. If the heights are not consistent, the reducer
re-emits the tuples that appeared early.
Since the incoming tuples are sorted by their position, it is enough
to use a counter to record how many tuples there are in each position
$i$. When the value of the counter becomes equal to the arity of the
conjunction, its means that all the subformulae hold at $i$ and the
reducer can emit the tuple for the conjunction at position $i$.
Otherwise, we reset the counter and continue.

\textbf{$\mathsf{U}_I$ modality.}
The reduce function for the \textit{Until} modality is
shown in Fig.~\ref{alg:reducer-until}.  When  we process tuples with
this function, we have to check both the temporal behavior and the
metric constraints (in the form of an $(a,b)$ interval) as defined by
the semantics of the modality.

Given a  formula $\phi_1\mathsf{U}_{(a,b)}\phi_2$, we check whether it can be evaluated 
in the current iteration, since reducer may receive some tuples
early. If this happens, reducer re-emits the tuple, as described above.

The algorithm processes each tuple $(\phi,i)$ sequentially. 
It keeps track of all the positions in the $(0,b)$ time window in the past with respect to the current tuple.
For each tuple it calls two auxiliary functions, \texttt{updateLTLBehavior} and \texttt{updateMTLBehavior}. 
The first function checks whether $\phi_1$ holds in all the positions
tracked in the $(0,b)$ time window; if this not the case 
we stop tracing these positions. This guarantee that we only keep
track of the position that exhibit the correct temporal semantics of
the \textit{Until} formula.  
Afterwards, function \texttt{updateMTLBehavior} checks the timing
constraints and removes positions that are outside of the $(0,b)$ time window.
Lastly, if $\phi_2$ holds in the position of the current tuple, we call 
function \texttt{emitUntil}, which emits an \textit{Until} tuple for
each position that we 
track, which is not in the $(0,a)$ time window in the past.

\textbf{$\mathfrak{C}$ modality.}
The reduce function for the $\mathfrak{C}$ modality is outlined in the
algorithm in Fig.~\ref{alg:reducer-count}.
To correctly determine if $\mathfrak{C}$ modality holds, we need to
keep track of all the positions in the past time window $(0,K)$.
While we sequentially process the tuples, we use variable $p$ to save
the position which appeared in the previous tuple. This allows us to
consider positions between each consecutive tuple in the inner ``for''
loop.
We call function \texttt{updateCountInterval}, which checks if the
tracked positions, together with the current one, occur within
the time
window $(0,K)$; positions that do not fall within the time interval
are discarded.
Variable $c$ is used to count   in how many tracked positions
subformula $\phi$ holds. 
At the end, we compare the value of $c$ with $n$ according to the
$\bowtie$ comparison operator; if this comparison is satisfied we emit
a $\mathfrak{C}$ tuple.

\textbf{$\mathfrak{M}$ modality.}
The algorithm in Fig.~\ref{alg:reducer-max} shows when the tuples for
the $\mathfrak{M}$ modality are emitted. Similarly to the
$\mathfrak{C}$ modality, we need to keep track of the all positions in
the $(0,K)$ time window in the past. Also, the two nested ``for''
loops make sure that we consider all time positions.  For each
position we call in sequence function \texttt{updateMaxInterval} and
function \texttt{emitMax}.  Function \texttt{updateMaxInterval} is
similar to \texttt{updateCountInterval}, i.e., it checks whether the
tracked positions, together with the current one, occur within the
time window $(0,K)$.  Function \texttt{emitMax} computes, in the
tracked positions, the maximum number of occurrences of the subformula
in all subintervals of length $h$.  It compares the computed value to
the bound $n$ using the $\bowtie$ comparison operator; if this
comparison is satisfied it emits the $\mathfrak{M}$ modality tuple.

\textbf{$\mathfrak{D}$ modality.}
The reduce function for the $\mathfrak{D}$ modality is shown in
Fig.~\ref{alg:reducer-dist}.  Similarly to the case of the
$\mathsf{U}_I$ modality, if the heights of the subformulae are not
consistent with the index of the current iteration, the reducer
re-emits the corresponding tuples.  After that, the incoming tuples
are processed in a sequential way and two nested ``for'' loops
guarantee that we consider all time points.  We need to keep track of all
the positions in the $(0,K)$ time window in the past in which either
$\phi$ or $\psi$ occurred. Differently from the previous aggregate
modalities, we have to consider only the occurrences of $\phi$ for which
there exists a matching occurrence $\psi$; for each of these pairs we
have to compute the distance. This processing of tuples (and the
corresponding atoms and time points that they include) is done by the
auxiliary function \texttt{updateDistInterval}.  Variables
$\mathit{pairs}$ and $\mathit{dist}$ keep track of the number of complete 
pairs in the current time window and their cumulative distance (computed
accessing the globally-shared map $\mathit{TS}$ of
timestamps).  Finally, by means of the function \texttt{emitDist}, if
there is any pair in the time window, we compare the average distance
computed as $\frac{dist}{pairs}$ with the bound $n$ using the
$\bowtie$ comparison operator. If the comparison is satisfied, we emit
a $\mathfrak{D}$ modality tuple.



\section{Related work}
\label{sec:rel}

To the best of our knowledge, the approach proposed
in~\cite{DBLP:conf/rv/BarreKSOH12} is the only one that uses \mr to
perform offline trace checking of temporal properties. The algorithm
is conceptually similar to ours as it performs iterations of \mr jobs
depending on the height of the formula.  However, the properties of
interest are expressed using LTL. This is only a subset of the
properties that can be expressed by \sol. Their implementation of the
conjunction and disjunction operators is limited to only two
subformulae which increases the height of the formula and results in
having more iterations. Intermediate tuples exchanged between mappers
and reducers are not sorted by the secondary key, therefore reducers
have to keep track of all the positions where the subformulae hold,
while our approach tracks only the data that lies in the relevant
interval of a metric temporal formula.

Distributed computing infrastructures and/or programming models have
also been used for other verification problems.
Reference~\cite{Lerda:1999:DMC:645879.672058} proposes a distributed
algorithm for performing \emph{model checking} of LTL \emph{safety
  properties} on a network of interconnected workstations.  By
restricting the verification to safety properties, authors can easily
parallelize a bread-first search algorithm.
Reference~\cite{monga13:dctl} proposes a parallel version of the
well-known fixed-point algorithm for CTL model checking. Given a set
of states where a certain formula holds and a transition relation of a
Kripke structure, the algorithm computes the set of states where the
superformula of a given formula holds though a series of \mr
iterations, parallelized over the different predecessors of the states
in the set.  The set is computed when a fixed-point of a predicate
transformer is reached as defined by the semantics of each specific
CTL modality.

\section{Evaluation}
\label{sec:eval}

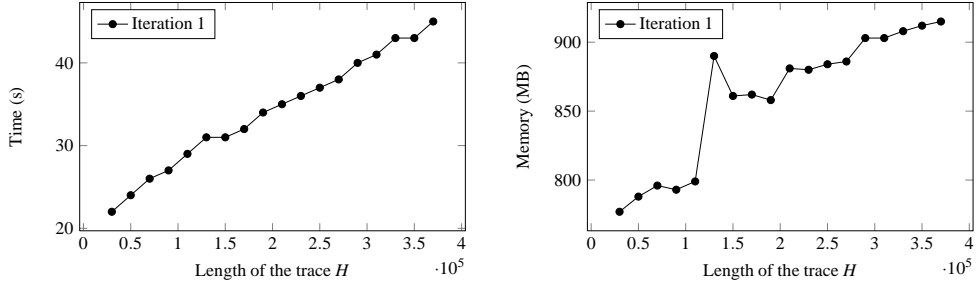
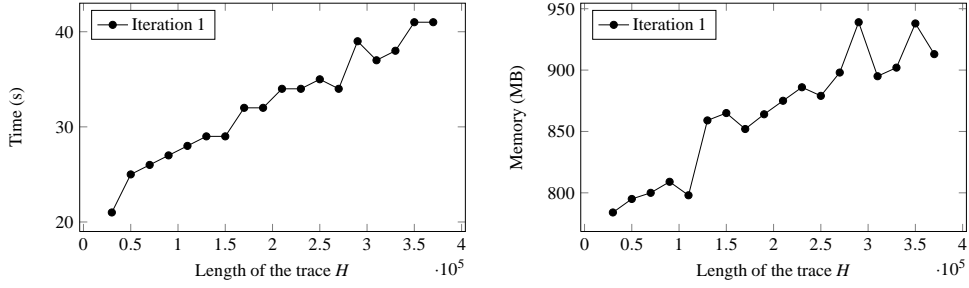
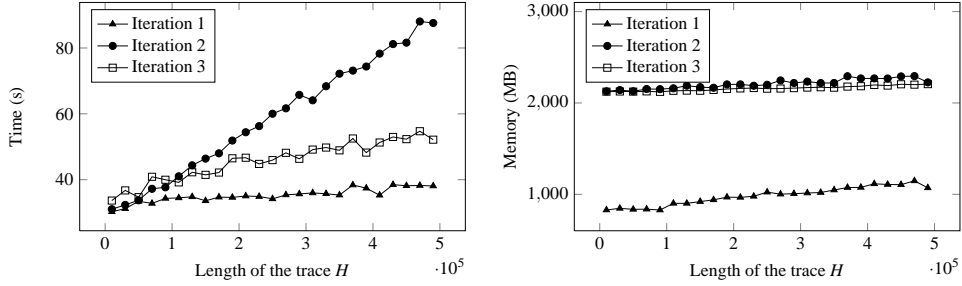
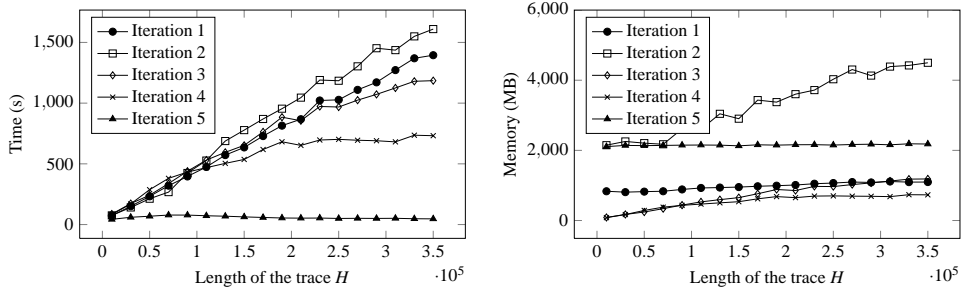
\begin{figure}
\centering
\begin{subfigure}{\textwidth}
\begin{tabular}{c c}
\begin{tikzpicture}[scale=\plotScale]
\begin{axis}[
     	height=\plotHeight,
     	width=0.7\textwidth,
		legend pos=north west,
		xlabel=Length of the trace $H$,
		ylabel=Time (s),	
		]


 \addplot[color=black, mark=*] table [x index=0, y index=7, col sep=comma] {csv/cmod-50k-0.csv};

\legend{Iteration 1}
\end{axis}
\end{tikzpicture} &
\begin{tikzpicture}[scale=\plotScale]
\begin{axis}[
		height=\plotHeight,
		width=0.7\textwidth,
		legend pos=north west,
		xlabel=Length of the trace $H$,
		ylabel=Memory (MB),
		]


 \addplot[color=black, mark=*] table [x index=0, y index=8, col sep=comma] {csv/cmod-50k-0.csv};

\legend{Iteration 1}
\end{axis}
\end{tikzpicture}\\
\end{tabular}
\caption{Formula: $\mathfrak{C}^{50000}_{< 10}(a_0)$}\label{fig:scal-small}
\end{subfigure}
\\
\vspace{0.3cm}
\begin{subfigure}{\textwidth}
\begin{tabular}{c c}
\begin{tikzpicture}[scale=\plotScale]
\begin{axis}[
		height=\plotHeight,
     	width=0.7\textwidth,
		legend pos=north west,
		xlabel=Length of the trace $H$,
		ylabel=Time (s),	
		]

\addplot[color=black, mark=*] table [x index=0, y index=7, col sep=comma] {csv/dist-20k-0.csv};

\legend{Iteration 1}
\end{axis}
\end{tikzpicture}&
\begin{tikzpicture}[scale=\plotScale]
\begin{axis}[
		height=\plotHeight,
     	width=0.7\textwidth,
		legend pos=north west,
		xlabel=Length of the trace $H$,
		ylabel=Memory (MB),
		]

\addplot[color=black, mark=*] table [x index=0, y index=8, col sep=comma] {csv/dist-20k-0.csv};

\legend{Iteration 1}
\end{axis}
\end{tikzpicture}\\
\end{tabular}
\caption{Formula: $\mathfrak{D}^{50000}_{< 10}(a_1,a_2)$}\label{fig:dist}
\end{subfigure}
\\
\vspace{0.3cm}
\begin{subfigure}{\textwidth}
\begin{tabular}{c c}
\begin{tikzpicture}[scale=\plotScale]
\begin{axis}[
		height=\plotHeight,
	    width=0.7\textwidth,
		legend pos=north west,
		xlabel=Length of the trace $H$,
		ylabel=Time (s),	
		]

\addplot[color=black, mark=triangle*] table [x index=0, y index=13, col sep=comma] {csv/scalabilityHigh3.csv};
\addplot[color=black, mark=*] table [x index=0, y index=13, col sep=comma] {csv/scalabilityHigh1.csv};
\addplot[color=black, mark=square] table [x index=0, y index=13, col sep=comma] {csv/scalabilityHigh2.csv};

\legend{Iteration 1,Iteration 2,Iteration 3}
\end{axis}
\end{tikzpicture}&
\begin{tikzpicture}[scale=\plotScale]
\begin{axis}[
		height=\plotHeight,
     	width=0.7\textwidth,
		legend pos=north west,
		xlabel=Length of the trace $H$,
		ylabel=Memory (MB),
		ymax = 3100
		]

\addplot[color=black, mark=triangle*] table [x index=0, y index=14, col sep=comma] {csv/scalabilityHigh1.csv};
\addplot[color=black, mark=*] table [x index=0, y index=14, col sep=comma] {csv/scalabilityHigh2.csv};
\addplot[color=black, mark=square] table [x index=0, y index=14, col sep=comma] {csv/scalabilityHigh3.csv};

\legend{Iteration 1,Iteration 2,Iteration 3}
\end{axis}
\end{tikzpicture}\\
\end{tabular}
\caption{Formula: $(a_0 \land (a_1 \land a_2))\mathsf{U}_{(50,200)}(( a_1 \land a_2 )\lor a_1)$}\label{fig:scal-big}
\end{subfigure}
\\
\vspace{0.3cm}
\begin{subfigure}{\textwidth}
\begin{tabular}{c c}
\begin{tikzpicture}[scale=\plotScale]
\begin{axis}[
		height=\plotHeight,
     	width=0.7\textwidth,
		legend pos=north west,
		xlabel=Length of the trace $H$,
		ylabel=Time (s),	
		]

\addplot[color=black, mark=*] table [x index=0, y index=13, col sep=comma] {csv/cmp4-1.csv};
\addplot[color=black, mark=square] table [x index=0, y index=13, col sep=comma] {csv/cmp4-2.csv};
\addplot[color=black, mark=diamond] table [x index=0, y index=13, col sep=comma] {csv/cmp4-3.csv};
\addplot[color=black, mark=x] table [x index=0, y index=13, col sep=comma] {csv/cmp4-4.csv};
\addplot[color=black, mark=triangle*] table [x index=0, y index=13, col sep=comma] {csv/cmp4-0.csv};

\legend{Iteration 1,Iteration 2, Iteration 3, Iteration 4, Iteration 5}
\end{axis}
\end{tikzpicture}&
\begin{tikzpicture}[scale=\plotScale]
\begin{axis}[
		height=\plotHeight,
     	width=0.7\textwidth,
		legend pos=north west,
		xlabel=Length of the trace $H$,
		ylabel=Memory (MB),
		ymax = 6000
		]

\addplot[color=black, mark=*] table [x index=0, y index=14, col sep=comma] {csv/cmp4-1.csv};
\addplot[color=black, mark=square] table [x index=0, y index=14, col sep=comma] {csv/cmp4-2.csv};
\addplot[color=black, mark=diamond] table [x index=0, y index=13, col sep=comma] {csv/cmp4-3.csv};
\addplot[color=black, mark=x] table [x index=0, y index=13, col sep=comma] {csv/cmp4-4.csv};
\addplot[color=black, mark=triangle*] table [x index=0, y index=14, col sep=comma] {csv/cmp4-0.csv};

\legend{Iteration 1,Iteration 2, Iteration 3, Iteration 4, Iteration 5}
\end{axis}
\end{tikzpicture}\\
\end{tabular}
\caption{Formula: $\exists j \in \{0\ldots 9\} \ \forall i \in \{0\ldots 8\}:\mathsf{G}_{(50,500)}(a_{i,j} \rightarrow \mathsf{X}_{(50,500)}(a_{i+1,j}))$}\label{fig:cmp4}
\end{subfigure}
\caption[]{Scalability of the algorithm}
\label{tab:scal}
\end{figure}

We have implemented 
the proposed trace checking  algorithm in Java using the Hadoop \mr
framework~\cite{foundation07:_hadoop_mapred} (version 1.2.1).
We executed it  on a Windows Azure cloud-based infrastructure where we allocated 10 small virtual machines with
1 CPU core and 1.75 GB of memory. We followed the standard Hadoop guidelines
when configuring the cluster: the number of map tasks was set to  the
number of nodes in the cluster multiplied by 10, and the number of
reducers was set to the number of nodes multiplied by 0.9; we used 100
mappers and 9 reducers. We have also enabled JVM reuse for any number of jobs, to minimize the time spent by framework in initializing Java virtual machines.
In the rest of this section, we first show 
 how the approach scales with respect to the trace length and 
how the height of the formula affects the running time and memory.
Afterwards, we compare our algorithm to the one presented in~\cite{DBLP:conf/rv/BarreKSOH12}, designed for LTL.

\begin{table}[t]
\scriptsize
\centering
\caption{Average processing time per tuple for the four properties.}
\label{tab:tupletime}
\begin{tabular}{c c c c c c c c c}
\toprule
&
\multicolumn{2}{c}{Property 1}&
\multicolumn{2}{c}{Property 2}&
\multicolumn{2}{c}{Property 3}&
\multicolumn{2}{c}{Property 4}\\

&
\sol&
LTL&
\sol&
LTL&
\sol&
LTL&
\sol&
LTL\\

\midrule

Number of tuples&
16,121& 55,009&
24,000& 119,871&
215,958& 599,425&
1,747,360 & 4,987,124\\

Time per event ($\mu$s)&
1.172& 19 &
1.894& 21 &
3.707& 14 &
7.200& 30\\

\bottomrule
\end{tabular}

\end{table}

\subsubsection*{Scalability.}
To evaluate scalability of the approach, we considered 4 formulae,
with different height: $\mathfrak{C}^{50000}_{< 10}(a_0)$,
$\mathfrak{D}^{50000}_{< 10}(a_1,a_2)$, $(a_0 \land (a_1 \land
a_2))\mathsf{U}_{(50,200)}(( a_1 \land a_2 )\lor a_1)$ and $\exists j
\in \{0\ldots 9\} \ \forall i \in \{0\ldots
8\}:\mathsf{G}_{(50,500)}(a_{i,j} \rightarrow
\mathsf{X}_{(50,500)}(a_{i+1,j}))$.  Here the $\forall$ and $\exists$
quantifiers are used as a shorthand notation to predicate on finite
domains: for example, $\forall i \in \{1,2,3\}: a_i$ is equivalent to
$a_1 \land a_2 \land a_3$.  We generated random traces with a number
of time instants varying from 10000 to 350000. For each time instant,
we randomly generated with a uniform distribution up to 100 distinct
events (i.e., \emph{atomic} propositions). Hence, we evaluated our
algorithm for a maximum number of events up to 35 millions.  The time
span between the first and the last timestamp was 578.7 days on
average, with a granularity of one second.

Figure~\ref{tab:scal} shows the total time and the memory used by the
\mr job run to check the four formulae on the generated traces. Formulae $\mathfrak{C}^{50000}_{< 10}(a_0)$ and 
$\mathfrak{D}^{50000}_{< 10}(a_1,a_2)$ needed one iteration
to be evaluated (shown in Fig.~\ref{fig:scal-small} and
Fig.~\ref{fig:dist}).
In both cases, the time 
taken to
check the formula increases linearly with respect to the trace
length; this happens because reducers need to process more
tuples. 
As for the linear increase in memory usage,  for modalities
$\mathfrak{C}$ and $\mathfrak{D}$ reducers have to keep track of all
the tuples in the window of length $K$ time units and the more time
points there are the more \emph{dense} the time window becomes, with a
consequent increase in memory usage.
As for the checking of the other two formulae (shown in Fig.~\ref{fig:scal-big} and
Fig.~\ref{fig:cmp4}), more iterations were needed because of the
height of the formulae. Also in this case, the time taken by each
iteration tends to increase as the length of the trace increases; the
memory usage is constant since the formulae considered here do not
contain aggregate modalities. Notice the increase of time and memory
from Fig.~\ref{fig:scal-big} to Fig.~\ref{fig:cmp4}: this is due to
the expansion of the quantifiers in formula $\exists j
\in \{0\ldots 9\} \ \forall i \in \{0\ldots
8\}:\mathsf{G}_{(50,500)}(a_{i,j} \rightarrow
\mathsf{X}_{(50,500)}(a_{i+1,j}))$.

\subsubsection*{Comparison with the LTL approach~\cite{DBLP:conf/rv/BarreKSOH12}.}
We compare our approach to the one presented
in~\cite{DBLP:conf/rv/BarreKSOH12}, which focuses on trace checking of
LTL properties using \mr; for this comparison we considered the LTL
layer included in \sol by means of the \textit{Until} modality.
Although the focus of our work was on implementing the semantics of
\sol aggregate modalities, we also introduces some improvements in the
LTL layer of \sol.  First, we exploited composite keys and secondary
sorting as provided by the \mr framework to reduce the memory used by
reducers. We also extended the binary $\land$ and $\lor$ operators to
support any positive arity.

We compared the two approaches by checking the following formulae:
\begin{inparaenum}[1)]
\item
$\mathsf{G}_{(50,500)}(\neg a_0)$;
\item $\mathsf{G}_{(50,500)}(a_0
\rightarrow \mathsf{X}_{(50,500)}(a_1))$;
\item $\forall i \in \{0\ldots
8\}:\mathsf{G}_{(50,500)}(a_{i} \rightarrow
\mathsf{X}_{(50,500)}(a_{i+1}))$; and
\item $\exists j \in \{0\ldots 9\}
\ \forall i \in \{0\ldots 8\}:\mathsf{G}_{(50,500)}(a_{i,j}
\rightarrow \mathsf{X}_{(50,500)}(a_{i+1,j}))$.
\end{inparaenum}
The height of these
formulae are 2, 3, 4 and 5, respectively.  This admittedly gives our
approach a significant advantage since
in~\cite{DBLP:conf/rv/BarreKSOH12} the restriction for the $\land$ and
$\lor$ operators to have an arity fixed to 2 results in a larger
height for formulae 3 and 4.
We randomly generated traces of variable length, ranging from 1000 to
100000 time instants, with up to 100 events per time instant. With
this configuration, a trace can contain potentially up to 10 million
events.  We chose to have up to 100 events per time instant to match
the configuration proposed in~\cite{DBLP:conf/rv/BarreKSOH12}, where
there are 10 parameters per formula that can take 10 possible values.
We generated 500 traces.  The time needed by our algorithm to check each of the four
formulae, averaged over the different traces, was 52.83, 85.38, 167.1
and 324.53 seconds, respectively. We do not report the time taken by
the approach proposed in~\cite{DBLP:conf/rv/BarreKSOH12} since the
article does not report any statistics from the run of an actual
implementation, but only metrics determined by a simulation.
  Table~\ref{tab:tupletime} shows the average number of tuples
 generated by the algorithm for each formulae. The number of tuples is
 calculated as the sum of all input tuples for mappers at each
 iterations in a single trace checking run. The table also shows the average time needed to process a single 
event in the trace. This time is computed as the total processing time
divided by the number of time instants in the trace, averaged over the
different trace checking runs. 
The \sol column refers to the data obtained by running our algorithm,
while the LTL column refers to data reported
in~\cite{DBLP:conf/rv/BarreKSOH12}, obtained with a simulation.
Our algorithm performs better both in terms of the number of generated
tuples and in terms of processing time.


\section{Conclusion and Future Work}
\label{sec:conc}

In this paper we present an algorithm based on the \mr programming
model that checks large execution traces against specifications
written in \sol.  The experimental results in terms of scalability and
comparison with the state of the art are encouraging and show that the
algorithm can be effectively applied in realistic settings.

A limitation of the algorithm is that reducers (that implement the semantics of 
temporal and aggregate operators) need to keep track of the positions relevant 
to the time window specified in the formula. In the future, we will 
investigate how this information may be split into smaller and more manageable parts
that may be processed separately, while preserving the original
semantics of the operators.

\subsubsection*{Acknowledgments.}
This work has been partially supported by  the National
Research Fund, Luxembourg (FNR/P10/03).

\appendix
\section{Auxiliary functions}
\label{sec:appx1}
In this section we present the pseudocode of the auxiliary functions
used in the reduce steps of \textit{until}, \textit{count}, \textit{maximum} and
\textit{distance} operators presented in Sect.~\ref{sec:reducer}.

As explained in Sect.~\ref{sec:reducer}, the reduce function for the \textit{until} 
modality of the form $\phi_1\mathsf{U}_{(a,b)}\phi_2$ keeps track of all the 
positions in the past $(0,b)$ time window with respect to the timestamp of the 
current tuple. 
For clarity, the positions are partitioned into arrays \texttt{int0A} and \texttt{intAB}
that store the positions in the past $(0,a]$ and $(a,b)$ time windows, respectively.
Function \texttt{clearIntervals}, shown in Fig.~\ref{alg:reducer-clear-interval}, removes all 
positions from both arrays.
Function \texttt{updateLTLBehavior}, shown in Fig.~\ref{alg:reducer-LTL-behaviour}, 
appends the current position to array \texttt{int0A} and checks whether the LTL 
condition of the \emph{until} operator holds for the stored positions. If the condition 
is violated, the arrays are cleared.
More specifically, the arrays are cleared if there exist some positions, 
between the current position $i$ and the previous position $p$, which were 
not received by the reducer. The arrays are also cleared if $\phi_1$ does not hold in all the 
consecutive positions currently stored by the reducer.
Function \texttt{updateMTLBehavior}, shown in Fig.~\ref{alg:reducer-MTL-behaviour}, 
is used to check the timing  conditions of the \emph{until} operator. Since \texttt{updateLTLBehavior} 
inserts the newly received position into the arrays, \texttt{updateMTLBehavior} updates 
the arrays with respect to the timestamp related to the new position.
If some positions from \texttt{int0A} are not in the $(0,a]$ interval
anymore, function \texttt{updateMTLBehavior} transfers them to
\texttt{intAB}. Next, the function removes all tuples from \texttt{intAB} that are not in the $(a,b)$ interval.
Function \texttt{emitUntil}, shown in Fig.~\ref{alg:reducer-until-emit}, emits \emph{until} tuples for
all positions stored in \texttt{intAB}.

The reduce function for the
$\mathfrak{C}^{K}_{\bowtie n}(\phi)$ modality keeps track of all 
the positions in the past time window $(0,K)$. The positions are stored in the 
array \texttt{intK}.
Function \texttt{updateCountInterval}, shown in Fig.~\ref{alg:reducer-count-interval}, 
adds the position from the current tuple to \texttt{intK} and then checks if all stored 
positions are in the past $(0,K)$ time window. In practice, the
function compares the difference between 
the timestamps of the first and the last position in \texttt{intK}. As
long as this difference 
is greater than $K$, the function removes the positions from the beginning of the array. It checks 
if the subformula $\phi$ holds at every removed position and, if it is
the case, it decrements 
variable $c$.

The reduce function for modality 
$\mathfrak{M}^{K,h}_{\bowtie n}(\phi)$ updates  its own corresponding 
array of positions \texttt{intK} using function
\texttt{updateMaxInterval}. It also computes, using \texttt{emitMax},
the maximum number of occurrences of subformula $\phi$
in subintervals of length $h$ over a window $K$.
Function \texttt{updateMaxInterval}, shown in Fig.~\ref{alg:reducer-max-interval},
checks whether the stored positions occur within the time window $(0,K)$ in 
the same way as its \textit{count} counterpart. 
Function \texttt{emitMax} calculates, for each position $z$ from \texttt{intK},
the subinterval $\mathsf{wc}$ it belongs to. This is done by calculating 
the difference between the timestamp at the last position in \texttt{intK} 
and the one at the position $z$ and then dividing it by the length of the 
subinterval $h$.
Variable $c$  counts the 
number of occurrences of subformula $\phi$ in the current 
subinterval $wc$.
We increment $c$
for every position where subformula $\phi$ holds and $\mathsf{wc}$ does not change 
with respect to the previous position. When $\mathsf{wc}$ changes, we update variable $\mathsf{max}$ and reset
variable $c$ to 0 or 1, depending on whether  subformula $\phi$ holds in the current position. 
When the function terminates, variable $\mathsf{max}$ holds the maximum number of 
$\phi$ occurrences in all subintervals. Finally, variable $\mathsf{max}$ is compared 
to the bound $n$ and a tuple is emitted in case the condition is satisfied.

The reduce function for the 
$\mathfrak{D}^{K}_{\bowtie n}(\phi,\psi)$ modality uses the \texttt{updateDistInterval} 
and \texttt{emitDist} functions, in a similar way as the previous modality.
Function \texttt{updateDistInterval}, shown in Fig.~\ref{alg:reducer-dist-interval}, updates the 
array of past positions \texttt{intK}. If subformula $\phi$ holds at a position that 
is removed from the array, we decrement the $\mathsf{pairs}$ variable that holds the current
number of $(\phi,\psi)$ pairs in \texttt{intK}. We also update the cumulative distance $\mathsf{dist}$ between complete pairs \texttt{intK}.
Finally, if there is at least one pair in the current time window, function \texttt{emitDist}
 (shown in Fig.~\ref{alg:reducer-dist-emit}) compares the average distance
computed as $\frac{dist}{pairs}$ to the bound $n$. If the condition is satisfied, it emits
a \textit{distance} modality tuple.

\setcounter{subfigure}{0}

\begin{figure}[!ht]
	\centering
	
	\begin{tabular}{cc}
	\begin{minipage}[t]{.5\textwidth}
	\centering
	\begin{algorithmic}
	\State Global variables: $a,b,int0A,intAB$
	\Function{updateMTLBehavior}{$i$}
		\State $\tau \gets TS(i)$
		\State $\tau_c \gets TS(\textit{int0A.first()})$
		\While{$\tau - a \geq \tau_c$}
			\State $p \gets \textit{int0A.removeFirst()}$
			\State $\textit{intAB.addLast(p)}$
			\State $\tau_c \gets TS(\textit{int0A.first()})$
		\EndWhile
		\State $\tau_c \gets TS(\textit{intAB.first()})$
		\While{$\tau - b \geq \tau_c$}
			\State $p \gets \textit{intAB.removeFirst()}$
			\State $\tau_c \gets TS(\textit{intAB.first()})$
		\EndWhile
	\EndFunction
	\end{algorithmic}{\tiny }

	\end{minipage} 
	&
	\begin{minipage}[t]{.5\textwidth}
	\centering
	\begin{algorithmic}
		\State Global variables: $p,\xi,\phi_1,int0A,intAB$
		\Function{updateLTLBehavior}{$i$}
		\If{$i - p > 1$}
			\State $\textit{clearIntervals()}$
		\EndIf
		\If{$\xi = \phi_1$}
			\If{$\textit{max(int0A.last(),intAB.last())!=i-1}$}
				\State $\textit{clearIntervals()}$
			\EndIf
			\State $\textit{int0A.addLast(i)}$
		\EndIf
		\EndFunction
	\end{algorithmic}{\tiny }
	\end{minipage} 
\\

	\begin{minipage}[t]{.5\textwidth}
		\captionof{subfigure}{Update MTL behavior function}\label{alg:reducer-MTL-behaviour}
	\end{minipage}&
	
	\begin{minipage}[t]{.5\textwidth}
		\captionof{subfigure}{Update LTL behavior function}\label{alg:reducer-LTL-behaviour}
	\end{minipage}\\

	\begin{minipage}[t]{.5\textwidth}
	\centering
	\begin{algorithmic}
	\State Global variables: $intK,\phi,c,K$
	\Function{updateCountInterval}{$i$}
		\State $\textit{intK.addLast(i)}$
		\While{$TS(intK.last)-TS(intK.first) > K$}
			\State $z \gets \textit{intK.removeFirst()}$
			\If{$(\phi,z) \in \textit{tuples}$}
				\State $c \gets c-1$
			\EndIf
		\EndWhile
	\EndFunction
	\end{algorithmic}{\tiny }

	\end{minipage} 
	&
	\begin{minipage}[t]{.5\textwidth}
	\centering
	\begin{algorithmic}
	\State Global variables: $intK,\phi,dist,pairs,K$
	\Function{updateDistInterval}{$i$}
		\State $\textit{intK.addLast(i)}$
		\While{$TS(intK.last)-TS(intK.first) > K$}
			\State $z \gets \textit{intK.removeFirst()}$
			\If{$z = subFmas.first()$}
				\State $subFmas.removeFirst()$
				\State $pairs \gets pairs-1$
				\State $dist \gets dist-(TS(subFmas.first-TS(z)))$
			\EndIf
		\EndWhile
	\EndFunction
	\end{algorithmic}{\tiny }
	\end{minipage}  
	\\
	
	\begin{minipage}[t]{.5\textwidth}
		\captionof{subfigure}{Update count interval function}\label{alg:reducer-count-interval}
	\end{minipage}&
	
	\begin{minipage}[t]{.5\textwidth}
		\captionof{subfigure}{Update distance interval function}\label{alg:reducer-dist-interval}
	\end{minipage}\\

	\begin{minipage}[t]{.5\textwidth}
 		\centering

 		\begin{algorithmic}
 		\State Global variables: $int0A,intAB$
 	 	\Function{clearIntervals}{\hspace{0mm}}
 		 	\State $\textit{intAB.clear()}$ 
 		 	\State $\textit{int0A.clear()}$
 	 	\EndFunction
 		\end{algorithmic}{\tiny }

 	\end{minipage} &

	\begin{minipage}[t]{.5\textwidth}
 		\centering

 		\begin{algorithmic}
 		\State Global variables: $intAB$
 	 	\Function{emitUntil}{$i$}
 	 	\ForAll{$z \in intAB$}
 		 	\State  output($\phi_1 \mathsf{U_{(a,b)}} \phi_2,z$)
 		 	\State  $\textit{intAB.remove(z)}$
 		 \EndFor
 	 	\EndFunction
 		\end{algorithmic}{\tiny }
 	\end{minipage} \\
 	
 	\begin{minipage}[t]{.5\textwidth}
 		\captionof{subfigure}{Clear intervals function}\label{alg:reducer-clear-interval}
 	\end{minipage}&
 	
 	\begin{minipage}[t]{.5\textwidth}
 		\captionof{subfigure}{Emit until tuples function}\label{alg:reducer-until-emit}
 	\end{minipage}\\
 	
	\multirow{3}{*}{
	\begin{minipage}[t]{.5\textwidth}
	\centering
	\begin{algorithmic}
	\State Global variables: $intK,\phi,K,n,h$
	\Function{emitMax}{$i$}
		\State $\tau_r \gets TS(intK.last)$
		\State $wc \gets 0$, $c \gets 0$, $max \gets 0$
		\ForAll{$z \in intK$}
			\If{$wc=\lfloor\frac{(\tau_r-TS(z))}{h}\rfloor$}
				\If{$(\phi,z) \in \textit{tuples}$}
					\State $c \gets c+1$
				\EndIf
			\Else
				\State $max \gets max(c,max)$
				\State $wc \gets (\tau_r-TS(z))/h$
				\If{$(\phi,z) \in \textit{tuples}$}
					\State $c \gets 1$
				\Else
					\State $c \gets 0$
				\EndIf
			\EndIf
		\EndFor
		\If{$max \bowtie n$}
			\State output($\mathfrak{M}_{\bowtie n}^{K,h}(\phi),i$)	
		\EndIf
	\EndFunction
	\end{algorithmic}{\tiny }
	\end{minipage}  
	}&

	\begin{minipage}[t]{.5\textwidth}
	\centering
	\begin{algorithmic}
	\State Global variables: $intK,K$
	\Function{updateMaxInterval}{$i$}
		\State $\textit{intK.addLast(i)}$
		\While{$TS(intK.last)-TS(intK.first) > K$}
			\State $\textit{intK.removeFirst()}$
		\EndWhile
	\EndFunction
	\end{algorithmic}{\tiny }
	\end{minipage}\\
	 	
	&
	\begin{minipage}[t]{.5\textwidth}
		\captionof{subfigure}{Update maximum interval function}\label{alg:reducer-max-interval}
	\end{minipage}\\
	 	 	
	&
	\begin{minipage}[t]{.5\textwidth}
		\centering
		\vspace{1cm}
		\begin{algorithmic}
		\State Global variables: $dist,pairs,n$
		\Function{emitDist}{$i$}
		\If{$pairs > 0$}
			\If{$\frac{dist}{pairs} \bowtie n$}
				\State output($\mathfrak{D}_{\bowtie n}^{K}(\phi,\psi),i$)	
			\EndIf
		\EndIf
		\EndFunction
		\end{algorithmic}{\tiny }
	\end{minipage}\\

	\begin{minipage}[t]{.5\textwidth}
		\captionof{subfigure}{Emit maximum tuples function}\label{alg:reducer-max-emit}
	\end{minipage}&
	
	\begin{minipage}[t]{.5\textwidth}
		\captionof{subfigure}{Emit distance tuples function}\label{alg:reducer-dist-emit}
	\end{minipage}\\

\end{tabular}
	
\caption{Auxiliary functions}\label{fig:aux}
\end{figure}


\end{document}